\documentclass[11pt]{article}
\usepackage[utf8]{inputenc}
\usepackage{amssymb}
\usepackage{amsmath}
\usepackage{amsfonts}
\usepackage{mathtools}
\usepackage{hyperref}
\usepackage{xcolor}
\usepackage{amsthm}

\hypersetup{
    colorlinks,
    linkcolor={blue!60!black},
    citecolor={blue!60!black},
    urlcolor={blue!60!black}
}

\usepackage{fullpage}

\usepackage{authblk}
\usepackage{xcolor}

\newcommand{\p}{\ensuremath{{\rm P}}}
\newcommand{\np}{\ensuremath{{\rm NP}}}
\newcommand{\fp}{\ensuremath{{\rm FP}}}
\newcommand{\conp}{\ensuremath{{\rm coNP}}}
\newcommand{\nl}{\ensuremath{{\rm NL}}}

\newcommand{\pspace}{\ensuremath{{\rm PSPACE}}}

\newcommand{\imp}{\ensuremath{\rightarrow}}
\newcommand{\ordimp}{\ensuremath{\mathord{\imp}}}
\newcommand{\absurd}{\ensuremath{\bot}}
\newcommand{\proves}{\ensuremath{\Rightarrow}}
\newcommand{\entails}{\ensuremath{\vdash}}
\newcommand{\datop}[2]{\genfrac{}{}{0pt}{0}{#1}{#2}}
\newcommand\discharge[2]{\stackrel{\mathclap{\tiny\mbox{(#2)}}}{#1}}
\newcommand\discharger[1]{\tiny\text{(#1)}}

\newcommand{\LG}{LG}
\newcommand{\LM}{\ensuremath{\text{LM}_{\ordimp}}}
\newcommand{\NM}{\ensuremath{\text{NM}_{\ordimp}}}

\newcommand{\mil}{minimal implicational logic}
\newcommand{\mpl}{minimal propositional logic}

\newtheorem{theorem}{Theorem}

\newtheorem{corollary}[theorem]{Corollary}

\title{A Closer Look at Some Recent Proof Compression-Related Claims\thanks{Supported in part by NSF grant CCF-2006496
.}}

\author{Michael C. Chavrimootoo}
\author{Ethan Ferland}
\author{Erin Gibson}
\author{Ashley H. Wilson}
\affil{Department of Computer Science\\University of Rochester\\Rochester, NY 14627, USA}

\date{December 23, 2022} 

\begin{document}

\maketitle

\begin{abstract}

Gordeev and Haeusler~\cite{gor-hae:j:compression-part-i} claim that each tautology $\rho$ of \mpl{} can be proved with a natural deduction of size polynomial in $|\rho|$. This builds on work from Hudelmaier~\cite{hud:j:nlogn-space-intui-logic} that found a similar result for intuitionistic propositional logic, but for which only the height of the proof was polynomially bounded, not the overall size. They arrive at this result by transforming a proof in Hudelmaier's sequent calculus into an equivalent tree-like proof in Prawitz's system of natural deduction, and then compressing the tree-like proof into an equivalent DAG-like proof in such a way that a polynomial bound on the height and foundation implies a polynomial bound on the overall size. 
Our paper, however, observes that this construction was performed only on \mil{}, which we show to be weaker than the \mpl{} for which they claim the result (see Section~\ref{s:imp-v-prop}).
Simply extending the logic systems used to cover \mpl{} would not be sufficient to recover the results of the paper, as it would entirely disrupt proofs of a number of the theorems that are critical to proving the main result.
Relying heavily on their aforementioned work, Gordeev and Haeusler~\cite{gor-hae:j:compression-part-ii} claim to establish $\np=\pspace$. The argument centrally depends on the polynomial bound on proof size in \mpl{}. Since we show that that bound has not been correctly established by them, their purported proof does not correctly establish $\np=\pspace$.
\end{abstract}

\section{Introduction}

This paper will discover and analyze flaws in two journal papers written by Gordeev and Haeusler~\cite{gor-hae:j:compression-part-i, gor-hae:j:compression-part-ii}. In particular, the two works in question are ``Proof Compression and NP versus PSPACE,'' which was published in \textit{Studia Logica} in 2019, and its follow-up, ``Proof Compression and NP versus PSPACE II,'' which was published in the \textit{Bulletin of the Section of Logic} in 2020. 

In their first paper, Gordeev and Haeusler~\cite{gor-hae:j:compression-part-i} consider Gentzen-style Sequent Calculus, denoted as SC, and Prawitz’s Natural Deduction, denoted as ND, in order to obtain the result that a formula $\rho$ is valid in minimal propositional logic if and only if there exists a DAG-like proof of $\rho$ whose size is polynomial in $\lvert\rho\rvert$. They arrive at this conclusion by first attempting to create an equivalent formal definition of minimal propositional logic, called \LM{}, which they attribute to the work of Hudelmaier~\cite{hud:j:nlogn-space-intui-logic}. Then, using a version of natural deduction, \NM{}, which is derived from Prawitz’s ND~\cite{pra:b:nat-deduction}, Gordeev and Haeusler~\cite{gor-hae:j:compression-part-i} argue that any valid formula $\rho$ has an \LM{} proof of length polynomial in the height and foundation of $\rho$. They go on to show that this \LM{} proof can be converted into an equivalent tree-like proof in \NM{} and can ultimately be horizontally compressed into a DAG-like proof with size polynomially bounded in $\lvert\rho\rvert$. The critical fault in this argument comes from Gordeev and Haeusler’s~\cite{gor-hae:j:compression-part-i} incorrect definition of minimal propositional logic. The purpose of the first part of this paper is to bring to light the issues that arise from the faulty definition. Further, in Section~\ref{s:error-rec}, this paper will address how possible extensions of the incorrect definition would still be insufficient to salvage Gordeev and Haeusler's~\cite{gor-hae:j:compression-part-i} result.

Gordeev and Haeusler's second paper~\cite{gor-hae:j:compression-part-ii}~is heavily reliant on the first paper. This subsequent paper calls attention to the problem of deciding whether a formula is provable in minimal logic, a problem that is $\pspace$-complete. Using this classification along with the result of their first paper, Gordeev and Haeusler claim to establish that $\np=\pspace$. Once again, there is a misconception about the equivalence of logical systems that causes significant damage to the argument. The second part of this critique will reveal the incorrect assumptions found in Gordeev and Haeusler's second paper, which are mostly attributable to the errors of the first paper. 

Note that this critique is largely focused on the first paper of Gordeev and Haeusler, but the importance of the critique is better appreciated in light of the latter paper. The result of $\np=\pspace$ would be highly influential in the field of complexity theory, having hierarchical collapsing repercussions such as and $\np=\conp$, $\p^{\#\p}=\np$, and $\np\neq\nl$. Thus, the considerations of this paper will mainly cover Gordeev and Haeusler's first work but hold the intentions of thwarting the propagation of undue results in the field of complexity theory.

\section{Preliminaries}

\subsection{Natural Deduction}\label{s:nat-ded}
Natural deduction is a way of formalizing languages of logic that involves creating proofs (also referred to as deductions) of logical sentences by application of a set of inference rules. These inference rules represent the very natural notion of deductive reasoning, where if some set of premises are known to be true, we can infer some conclusion. The rules are defined using what can be thought of as templates of logical sentences. Instead of using individual propositional variables, the templates use variables that represent entire sentences of their own. For example, the template $A\lor B$ could represent infinitely many sentences including $p\lor q$, $(p\land q)\lor(r\imp q)$, etc.
The inference rules are defined with the following notation
\[ \frac{\alpha_1 \quad \alpha_2 \quad\cdots\quad \alpha_n}{\beta}, \]
where $\beta$ can be immediately derived from the $\alpha_i$s, i.e., if every $\alpha_i$ is known to be true, then $\beta$ is also known to be true.
Application of these rules to derive some sentence $\rho$ results in a tree for which $\rho$ is the root, and each parent is related to its children by a particular inference rule. Each leaf of the tree is an assumption. Anything can be assumed, but the derived sentence only holds under the assumptions (unless they are discharged---see below).

For example, suppose we have a system of natural deduction that has the rules $\frac{A\ B}{A\land B}$ and $\frac{A\land B}{A}$. If we wanted to derive $(p\land q)\land r$ from the assumptions $p$, $q$, and $r\land s$, we could do so with the following derivation:
\[
\dfrac{\dfrac{p\quad q}{p\land q}\quad\dfrac{r\land s}{r}}{(p\land q)\land r}.
\]

With the system as is, sentences can only be derived under a nonempty set of assumptions, so there are no tautologies (sentences that are always true; they can be derived without assumptions). This is where the notion of discharging assumptions comes in. Inference rules can be defined so that their application discharges specific assumptions, meaning that those assumptions are accounted for and are no longer assumed. This is indicated in the inference rule by writing the assumption being discharged in parentheses above the $\alpha$ for which it applies. We can expand the formalization of the notation for inference rules to
\[ \frac{\datop{(\gamma_1)}{\alpha_1} \datop{\ldots}{\ldots} \datop{(\gamma_k)}{\alpha_k} \quad \datop{}{\alpha_{k+1}} \datop{}{\ldots} \datop{}{\alpha_n}}{\beta}, \]

where, for each $i \in \{1, \ldots, k\}$, whenever $\gamma_i$ is an assumption of the subderivation leading to $\alpha_i$, $\gamma_i$ is discharged. In the derivation, every rule application that discharges an assumption is labeled with a number, and that number is written above every instance of the assumption it discharges.

Returning to the earlier example, suppose we add this new inference rule that allows us to discharge any assumption:
\[\dfrac{\datop{(A)}{B}}{A\imp B}.\]
We can now prove the tautology $(r\land s)\imp(q\imp(p\imp((p\land q)\land r)))$ by extending the derivation from above and discharging all the assumptions.
\[
\dfrac{\dfrac{\dfrac{\dfrac{\dfrac{
\discharge{p}{1}\quad \discharge{q}{2}}{
p\land q}\quad\dfrac{
\discharge{r\land s}{3}}{r}}{
(p\land q)\land r}}{
p\imp((p\land q)\land r)}\discharger{1}}{
q\imp(p\imp((p\land q)\land r))}\discharger{2}}{
(r\land s)\imp(q\imp(p\imp((p\land q)\land r)))}\discharger{3}.
\]
We can clearly see that this is a tautology because every leaf in the tree has a number above it, so every assumption is discharged.

For a more detailed exposition of natural deduction, we refer readers to Prawitz's monograph on the topic~\cite{pra:b:nat-deduction}.

\subsection{Sequent Calculus}
Sequent calculus is another approach to logical derivation. Compared to natural deduction, it has a similar notation for inference rules and a similar tree-like proof structure, but differs greatly in how it handles assumptions. Rather than single logical sentences, each node in the deduction (an $\alpha$ or $\beta$ in the notation specified in Section~\ref{s:nat-ded}) is a sequent---a conditional tautology expressed as $\Gamma \proves \Delta$ where $\Gamma$ is a set of antecedents and $\Delta$ is a single consequent.\footnote{
In some systems, $\Delta$ could be set of consequents rather than a singe consequent, but we will not consider such systems here.}
The consequent is what can be proved given that all of the antecedents are true. The sequent essentially bundles the assumptions together with the consequent, making each sequent in some sense self-contained: while the full derivation is needed to establish validity of the sequent, each individual sequent in the derivation is true on its own; there are no separate assumptions in other parts of the derivation that must be accounted for. This removes the need to discharge assumptions, and instead we can simply remove antecedents. An unconditional tautology would be any consequent for which there are no antecedents. In other words, any $\Delta$ for which $\Gamma$ is the empty set.
Sequent calculi include axioms in addition to their inference rules. Rather than each leaf of the deduction tree being an assumption, each leaf must be an axiom.
In both axioms and inference rules, a subset of the antecedents can be specified while the rest of the antecedents are left free using the notation $\Gamma,\alpha_1\ldots\alpha_n$.

As an example, consider a sequent calculus with just the axiom $\Gamma,p \proves p$ and the inference rule $\frac{\Gamma, A\proves B}{\Gamma \proves A\imp B}$.\footnote{This inference rule is essentially the sequent calculus analogue of the last natural deduction rule from the example in Section~\ref{s:nat-ded} that was for discharging assumptions.} We could use this system to prove the unconditional tautology $p\imp (q \imp p)$ as follows:
\[
\dfrac{\dfrac{q,p \proves p}{p \proves q\imp p}}{\proves p\imp(q\imp p)}.
\]

For a more detailed exposition of sequent calculus, we refer readers to Appendix~A of Prawitz~\cite{pra:b:nat-deduction}.

\subsection{Minimal and Intuitionistic Logic}

Intuitionistic logic is a logic which rejects both the law of excluded middle ($\entails p \lor \lnot p$) and double negation elimination ($\lnot\lnot p\entails p$). Minimal logic, originally introduced by Johansson~\cite{joh:j:minimal-logic}, is an intuitionistic logic that additionally rejects the principle of explosion ($\absurd\entails p$).

The notions of intuitionistic and minimal logic can be applied to different syntaxes of logic. They are traditionally formalized in the context of predicate logic~\cite{joh:j:minimal-logic,pra:b:nat-deduction} but may also be applied in the context of propositional logic (as in~\cite{hud:j:nlogn-space-intui-logic}) or even purely implicational logic (discussed in~\cite{pog:b:implicational-logic-chapter}). Regardless of the syntactic context, the restrictions relative to classical logic remain the same, and the differences between, for example, intuitionistic propositional logic and intuitionistic predicate logic are the same as the differences between classical propositional logic and classical predicate logic (namely the inclusion of quantifiers, quantified variables, and relations). 

It is worth noting that negation ($\lnot$) is generally not included as a propositional connective in intuitionistic or minimal logic. Rather, absurdity ($\absurd$) is added as a nullary connective, where $\lnot p$ is interpreted as a shorthand for $p\imp \absurd$.

Prawitz formalizes systems of natural deduction for minimal, intuitionistic, and classical logic~\cite{pra:b:nat-deduction}. The system for minimal logic uses the following inference rules:
\begin{align*}
    \land I: &\quad \frac{A\quad B}{A\land B} &
    \lor I: &\quad \frac{A}{A\lor B} \quad \frac{B}{A\lor B} &
    \ordimp I: & \quad \frac{\datop{(A)}{B}}{A\imp B} \\
    \\
    \land E: &\quad \frac{A\land B}{A}\quad \frac{A\land B}{B} &
    \lor E: &\quad \frac{\datop{}{A\lor B} \  \datop{(A)}{C} \ \datop{(B)}{C}}{C} &
    \ordimp E: &\quad \frac{A\quad A\imp B}{B}. \\
\end{align*}
These rules consist of one introduction rule and one elimination rule for each standard (binary) propositional connective. Prawitz also included introduction and elimination rules for $\forall$ and $\exists$ because the system was formulated in predicate logic, but we can simply take the propositional fragment (i.e., remove the rules for quantifiers) to get a system for \mpl{}.

Prawitz's system for intuitionistic logic includes all of the above inference rules along with a rule for the principle of explosion, formalized as:
\[ \absurd_I:\quad \frac{\absurd}{A}. \]

\subsection{Hudelmaier's Sequent Calculus}
Hudelmaier presents a $\pspace$ algorithm for intuitionistic propositional logic~\cite{hud:j:nlogn-space-intui-logic}. The procedure uses backward application of the rules of a sequent calculus (referred to as \LG{}). The $\pspace$ classification is achieved because \LG{} was constructed to have the critical property that every deduction of a sequent $s$ is linearly bounded in length with respect to the length of $s$. This is not a linear (or even polynomial) bound on the overall size of the deduction, only the length (as some of the rules produce multiple branches).

The calculus \LG{} includes the axioms $\Gamma,p \proves p$ and $\Gamma,\absurd \proves p$. The latter axiom is the principle of explosion, and thus can be excluded to create a \mpl{} version of the calculus. \LG{} uses 12 inference rules labeled: $GI1\land$, $GI2\land$, $GI1\lor$, $GI2\lor$, $GI1\ordimp$, $GI2\ordimp$, $GE\land$, $GE\lor$, $GE\ordimp P$, $GE\ordimp\land$, $GE\ordimp\lor$, $GE\ordimp\ordimp$. For brevity, these will not be listed here in their full form, but they use the following naming conventions: Rules in the form $GIXC$, where $X\in \{1,2\}$ and $C$ is a propositional connective, are introduction rules that introduce $C$ in the consequent. Rules in the form $GEC$ are elimination rules that introduce $C$ in the antecedent. The four $GE\ordimp$ rules introduce an implication in the antecedent, in which the antecedent of said implication includes the specified connective or, in the case of $GE\ordimp P$ is just a propositional variable.

\subsection{Computational Complexity}
In this paper, we assume familiarity with basic concepts of complexity theory. Readers may wish to consult any standard
text on the subject, e.g.,~\cite{pap:b:complexity,sip:b:introduction-third-edition}.

The complexity class $\np$ is the class of languages that can be decided by a nondeterministic Turing machine in polynomial time. The class $\pspace$ is the class of languages that can be decided by a deterministic Turing machine using polynomial space. The class $\fp$ is the
set of polynomial-time computable functions.
It is well-known that $\np \subseteq \pspace$, and whether $\np = \pspace$ is a central open issue in theoretical computer science.

Formally, given two sets (i.e., decision problems) $A$ and $B$, we say
that $A \leq_m^p B\iff (\exists f \in \fp)(\forall x)[x \in A \iff f(x) \in B]$, i.e., $A$ (polynomial-time) many-one reduces to $B$. 
Many-one reductions are tremendously important in complexity theory, as
they allow one to relate the hardness of one problem to that of another.
One such way is via ``complete problems.''

Formally, a $\pspace$-complete problem $B$ is a $\pspace$ problem
such that $(\forall A \in \pspace)[A \leq_m^p B]$. Informally,
$\pspace$-complete problems are the ``hardest'' problems in $\pspace$.
It is well-known that polynomial-time many-one reductions are transitive
and that $\np$ is closed downwards under polynomial-time many-one reductions (i.e., if $A \leq_m^p B$ and $B \in \np$, then $A \in\np$).
It follows that if some $\pspace$-complete problem is shown to be in $\np$, then $\np=\pspace$.

\section{Overview of the Polynomial Proof-Size Argument of~\protect\cite{gor-hae:j:compression-part-i}}
Gordeev and Haeusler~\cite{gor-hae:j:compression-part-i} introduce a positive implicational subsystem of Hudelmaier's \LG{} calculus~\cite{hud:j:nlogn-space-intui-logic}, which they call \LM{}. They claim~\cite[Claim~1]{gor-hae:j:compression-part-i} that \LM{} is sound and complete for \mpl{}, and that it inherits the properties of \LG{}~\cite[Lemma~2.2-3]{gor-hae:j:compression-part-i}. They further claim that deductions in \LM{} have the ``semi-subformula property,'' meaning that any formula occurring in an \LM{} deduction of $\rho$ is either a proper subformula of $\rho$ or in the form $\beta\imp \gamma$ for some semi-subformula $(\alpha\imp \beta)\imp\gamma$ of $\rho$~\cite[Lemma~2.1]{gor-hae:j:compression-part-i}. A proof of an upper bound on the number of semi-subformulas establishes that the foundation (number of distinct formulas occurring in a deduction) of any tree-like deduction of $\rho$ in \LM{} is polynomial in $|\rho|$ (the length of the formula)~\cite[Lemma~2.4]{gor-hae:j:compression-part-i}.

\LM{} is a sequent calculus, but the compression technique that will be used only works with natural deduction, so the authors introduce \NM{}: the implicational fragment of Prawitz's ND system for minimal logic~\cite{pra:b:nat-deduction}. They modify it to include some repetition rules and claim soundness and completeness with respect to \mpl{}~\cite[Claim~3, Lemma~5]{gor-hae:j:compression-part-i}. They then present a recursive transformation from any derivation in \LM{} to \NM{}, such that the height of the \NM{} derivation is polynomial with respect to the height of the \LM{} derivation, and therefore polynomial in $|\rho|$~\cite[Theorem~4]{gor-hae:j:compression-part-i}.

While the height and foundation of the tree-like deduction have polynomial bounds, the overall size could still be exponential due to repetition of identical formulas in different branches of the tree. To get around this, the Gordeev and Haeusler introduce the notion of DAG-like deductions~\cite[Section~2]{gor-hae:j:compression-part-i}. This allows multiple parents to have the same child, so proofs of a semi-subformula $\alpha$ need not be repeated if $\alpha$ is needed in multiple places.

Next, they outline a compression algorithm to take a tree-like deduction in \NM{} and convert it into a DAG-like deduction~\cite[Section~3]{gor-hae:j:compression-part-i}. DAG-like deductions use a directed acyclic graph instead of a tree to represent the derivations, allowing repeated sub-derivations to be merged into one single sub-derivation that is referenced in multiple different places in the derivation. This leads to the result that any tautology of \NM{} has a DAG-like deduction with overall size polynomial in $|\rho|$~\cite[Theorem~14]{gor-hae:j:compression-part-i}. Using the soundness and completeness results from earlier, they conclude that any tautology of \mpl{} has a DAG-like proof in \NM{} with size polynomial in $|\rho|$~\cite[Corollary~15]{gor-hae:j:compression-part-i}. This only establishes that tautologies are provable by small DAG-like deductions. Finally, they prove the other direction: that all small DAG-like deductions in \NM{} do represent tautologies~\cite[Theorem~19]{gor-hae:j:compression-part-i}. Together this gives the soundness and completeness with respect to \mpl{} of DAG-like \NM{} provability~\cite[Corollary~20]{gor-hae:j:compression-part-i} and the most central result (which is Corollary~21 in the paper):
\begin{corollary}[\cite{gor-hae:j:compression-part-i}]\label{c:small-mpl}
    A given formula $\rho$ is valid in \mpl{} iff there exists a DAG-like \NM{} proof of $\rho$ whose size is $O(|\rho|^4)$
\end{corollary}

The paper ends by addressing the complexity of verifying small DAG-like proofs in \NM{}, but does not reach a definitive conclusion~\cite[Section~5]{gor-hae:j:compression-part-i}.

\section{Errors in the Polynomial Proof-Size Argument of~\protect\cite{gor-hae:j:compression-part-i}}\label{s:error}

The argument that builds to Corollary~\ref{c:small-mpl}~\cite{gor-hae:j:compression-part-i} fails due to a conflation of \mil{} and \mpl{}, which are not equivalent.
While both are types of minimal logic, they have a different syntax. Specifically, \mil{} is a fragment of \mpl{}, meaning that \mil{} is a subset of \mpl{} formed by restricting the syntax. While \mpl{} allows all standard propositional connectives ($\land$, $\lor$, $\ordimp$, $\absurd$), \mil{} allows only implication ($\ordimp$) (and sometimes $\absurd$, depending on the formalization).

The authors note that they are considering a language of minimal logic using only propositional variables and the connective $\imp$~\cite[Section~1]{gor-hae:j:compression-part-i}. This is \mil{}, but the authors seem to be under the impression that it is equivalent to \mpl{}. Rather than take the difference in logic systems for granted and just assume the results don't scale, this section will firmly establish why their proof construction and compression does not work for \mpl{}.

\subsection{Omissions from Referenced Proof Systems}
The two calculi introduced in the paper, \LM{} and \NM{}, are not novel systems---they are modifications to existing systems proposed by Hudelmaier~\cite{hud:j:nlogn-space-intui-logic} and Prawitz~\cite{pra:b:nat-deduction}, respectively. Thus, proofs of their soundness and completeness with respect to \mpl{} are delegated to the papers from which the systems originated~\cite[Claim~1, Claim~3]{gor-hae:j:compression-part-i}. However, the modifications are substantial and equivalence is not proved, so delegating the proofs is not a valid strategy. Thus, \LM{} and \NM{} are not established to be sound and complete for \mpl{}.

In both cases, all rules that are not purely implicational are left out, as well as all rules including $\absurd$. In converting from \LG{}~\cite{hud:j:nlogn-space-intui-logic} to \LM{}, this includes the second axiom ($\Gamma,\absurd \proves p$), which is correctly omitted to modify the intuitionistic logic to a minimal logic, along with $GI1\land$, $GI2\land$, $GI1\lor$, $GI2\lor$, $GE\land$, $GE\lor$, $GE\ordimp\land$, and $GE\ordimp\lor$, which are incorrectly omitted, resulting in a modification from \mpl{} to \mil{}. In converting from Prawitz's minimal logic~\cite{pra:b:nat-deduction} to \NM{}, rules $\land I$, $\land E$, $\lor I$, and $\lor E$ are incorrectly omitted, again resulting in a modification from \mpl{} to \mil{}. These inference rules are needed to prove many things in \mpl{}, so omitting them weakens the system.

\subsection{Incompleteness of \texorpdfstring{$\LM$}{LM\_{→}} and \texorpdfstring{$\NM$}{NM\_{→}}}\label{s:imp-v-prop}
As a counterexample to the completeness of \LM{} and \NM{} with respect to \mpl{}, consider
\[(p\land q)\imp p.\]
This is clearly provable in \mpl{}, as demonstrated by the following natural deduction proof:
\begin{align*}
\dfrac{\dfrac{\discharge{p \land q}{1}}{p}}{(p \land q) \imp p}\discharger{1} \quad
\datop{(\land E)}{(\ordimp I)}.
\end{align*}

Pure implication is not functionally complete and thus cannot express all the standard propositional connectives. \LM{} and \NM{}, being purely implicational systems, cannot directly express, for example, $\land$. This alone could be considered sufficient evidence against the completeness of \LM{} and \NM{}; if they cannot even express certain tautologies of \mpl{} they have no hope of proving the tautologies. However, one might argue that a system that is able to prove the same notions in a different notation could still be considered complete for \mpl{}. At the very least, the overall results of the paper would hold if \mpl{} could be polynomial-time many-one reduced to \mil{}.

The functional incompleteness of implication can be resolved with the addition of $\absurd$. For example, $p\land q$ could be expressed as $(p \imp (q \imp \absurd))\imp \absurd$. Since each standard propositional connective could be replaced with an equivalent expression using only $\imp$ and $\absurd$ in polynomial time, we can permit this transformation with regards to time complexity results. The paper does establish that the language of minimal logic being used does not include $\absurd$~\cite[Section~1]{gor-hae:j:compression-part-i}, but since allowing $\absurd$ to take the place of any predicate in an existing axiom or rule does not impact any of their proofs, we will permit it in an attempt to recover the results of the paper.

Using only implication and $\absurd$, $(p\land q)\imp p$ can be expressed as:
\[((p \imp (q \imp \absurd)) \imp \absurd)\imp p.\]
There are other ways it could be expressed, but this is the simplest and most standard. The below argument will work in exactly the same way for any alternative choice of how to represent $\land$.\footnote{It is worth noting that the representation must treat the $(p\land q)$ clause as atomic, and thus cannot restructure the entire sentence to something like $p\imp(q\imp p)$. While $p\imp(q\imp p)$ also expresses a similar idea, it cannot be used for a transformation from \mpl{} to \mil{} because it depends on the $\imp$ following the $\land$ clause, and thus cannot be generalized.}

Because the proof systems \LM{} and \NM{} have no axioms or inference rules specifically including $\absurd$, formulas including $\absurd$ can only be derived by treating $\absurd$ as an ordinary subformula, no different from some arbitrary $A$. In other words, the special meaning we have assigned to $\absurd$ as a nullary connective cannot be leveraged in derivations. Any derivation involving $\absurd$ could be modified by replacing every occurrence of $\absurd$ with some new propositional variable, and each rule application would remain valid, as $p$ could take the place of a variable in an inference rule just as easily as $\absurd$ could. For the same reason, any propositional variable could be replaced with $\absurd$, as the variable cannot be broken down 
into smaller components to take advantage of some rule that $\absurd$ cannot. Essentially, $\absurd$ and $p$ are both atomic and indistinguishable from the perspective of the inference rules in \LM{} and \NM{}, so they are interchangeable. For example, $\absurd\imp (p \imp \absurd)$ is a theorem in \LM{} and \NM{} precisely because $q\imp (p \imp q)$ is also a theorem in \LM{} and \NM{}.

Thus, $((p \imp (q \imp \absurd)) \imp \absurd)\imp p$ is a theorem in \LM{} and \NM{} if and only if $((p \imp (q \imp r)) \imp r)\imp p$ is a theorem. The latter is obviously not a tautology in \mpl{} (or even in the more permissive classical logic---consider the case where $p$ is false and $r$ is true), and therefore is not a theorem either since \LM{} and \NM{} are both sound. 

We can thus conclude that $(p\land q)\imp p$ is a tautology in \mpl{} but is not a theorem of \LM{} or \NM{}, and cannot be transformed into an equivalent implicational theorem of \LM{} or \NM{}. Therefore, \LM{} and \NM{} are not complete with respect to \mpl{}.

\subsection{Ramifications of the Errors}
The paper claims that tree-like provability in $\NM{}$ is sound and complete with respect to \mpl{}~\cite[Lemma~5]{gor-hae:j:compression-part-i}. We have established this to be false. It makes the same claim for DAG-like provability in $\NM{}$~\cite[Corollary~20]{gor-hae:j:compression-part-i}, but depends on the soundness and completeness of tree-like provability~\cite[Lemma~5]{gor-hae:j:compression-part-i} to do so, so it does not hold. Corollary~\ref{c:small-mpl}~\cite[Corollary~21]{gor-hae:j:compression-part-i}, the central result of the paper, depends on soundness and completeness of DAG-like provability~\cite[Corollary~20]{gor-hae:j:compression-part-i}, and thus it also does not hold. Therefore, the paper has not proven that all (and only) tautologies of \mpl{} have a proof whose size is polynomial in the size of the tautology.

\subsection{Challenges in Recovering from the Errors}\label{s:error-rec}
One might try to recover the results of the paper by extending \LM{} and \NM{} to include the omitted rules and updating the transformation and compression algorithms to account for $\land$ and $\lor$. This would require adding at least eight additional cases to the recursive function that converts from \LM{} to \NM{} to handle the eight missing rules of \LG{}, and would disrupt many of the assumptions and proofs throughout the paper.

Notably, the inclusion of the rest of the rules from \LG{} would invalidate the semi-subformula property~\cite[Lemma~2.1]{gor-hae:j:compression-part-i}. For a counterexample, consider inference rule $GE\ordimp\lor$:\footnote{
The notation has been slightly modified from the original paper to maintain consistency of notation within this paper; the semantics remain identical. Note that $p$ is a propositional variable not occurring in $A$, $B$, $C$, or $D$.
}
\[
\frac{\Gamma,A\imp p, B\imp p, p\imp C \proves D}{\Gamma,(A\lor B)\imp C\proves D}.
\]
This involves multiple formulas that cannot appear in the resulting formula that they are being used to prove. Thus, even a weaker notion of the semi-subformula property (one that allows more nonproper subformulas beyond just $\beta\imp\gamma$ as a semi-subformula of $(\alpha\imp\beta)\imp\gamma$) could not hold, as arbitrary propositional variables can appear in a deduction in \LG{} that are not anywhere in the formula being proved.

The semi-subformula property is crucial for proving the polynomial bound on foundation in \LM{}~\cite[Lemma~2.4]{gor-hae:j:compression-part-i}. The bound on foundation depends on a polynomial bound on the number of distinct semi-subformulas of any given formula, but if the semi-subformula property is lost and arbitrary formulas can appear in the deduction that are not semi-subformulas, the bound on foundation is lost as well. Without a polynomial bound on foundation in \LM{}, the polynomial bound on foundation for deductions in \NM{}~\cite[Theorem~4]{gor-hae:j:compression-part-i} would also fail to hold, as it relies on property carryover from \LG{}. The bound on foundation for \NM{} is what allows the compression algorithm to get a polynomial bound in size. Only identical formulas can be merged in the DAG, so if there is no polynomial bound on foundation then the compression cannot get a polynomial bound on size~\cite[Corollary~15]{gor-hae:j:compression-part-i}. Without the polynomial bound on size, Corollary~\ref{c:small-mpl}~\cite[Corollary~21]{gor-hae:j:compression-part-i} is similarly lost. Thus even the substantial additions that would be required to cover the rest of \mpl{} are not sufficient to recover the central results of the paper.

\section{Implications for the Claimed Resolution of \texorpdfstring{\boldmath$\np$}{NP} vs. \texorpdfstring{$\pspace$}{PSPACE}}

In a subsequent paper,~\cite{gor-hae:j:compression-part-ii},
Gordeev and Haeusler build on their earlier work~\cite{gor-hae:j:compression-part-i} and
claim to prove that $\np=\pspace$.
In their proof, they claim that the $\pspace$-complete problem
of deciding if a formula is provable in minimal logic is also in $\np$
by leveraging
their polynomial upper bound on the size of proofs in minimal propositional
logic~\cite{gor-hae:j:compression-part-i}. However, as we saw in the previous section, 
that proof is flawed and does not actually provide a polynomial upper bound
on the size of proofs in minimal propositional logic. In this section, we
briefly detail some
of the errors made by Gordeev and Haeusler~\cite{gor-hae:j:compression-part-ii} and conclude
that the paper fails to establish that $\np$ equals $\pspace$.

We first note that the paper cites Statman~\cite{sta:j:intui-logic}
and Svejdar~\cite{sve:j:intui-logic} as sources
for the $\pspace$-completeness of the problem of deciding if a formula
is provable in minimal logic (in both papers by Gordeev and 
Haeusler~\cite{gor-hae:j:compression-part-i,gor-hae:j:compression-part-ii}, minimal logic is actually minimal implicational logic, although they incorrectly call it \mpl). However, upon closer inspection, these
papers give completeness results for the problem of deciding if a formula
is provable in \emph{intuitionistic} propositional logic, which is a stronger form of logic than minimal implicational logic. We are not aware of 
analogous $\pspace$-complete problems with respect to minimal propositional
logic. However, for the sake of argument, we entertain the notion that
such results may exist and try to see where the argument made 
in~\cite{gor-hae:j:compression-part-ii} fails.

\begin{theorem}[\cite{gor-hae:j:compression-part-ii}]\label{t:equal}
$\pspace\subseteq \np$ and hence $\np = \pspace$.
\end{theorem}

In their proof of Theorem~\ref{t:equal} (which is Theorem~2.8 in the
original paper), the authors argue that the theorem follows directly from
\cite[Corollary~15]{gor-hae:j:compression-part-i}. Fortunately, they also
explain how it follows, and from that we can easily spot the error, thanks
to our observations from the previous section. Their argument starts as follows:
\begin{quote}
    First we ``guess'' the existence of ``short'' 
    Hudelmaier-style cutfree sequential deduction
of $\rho$ that leads (by deterministic compression) to a ``small'' natural
deduction frame $\tilde{D}$ that is supposed to have \textit{fst} $\mathcal{F}$. 
Then we ``guess'' the existence of a ``cleansed'' modified subdeduction that confirms
in $|\rho|$-polynomial time the provability of $\rho$ with regard to $\langle \tilde{D}, \mathcal{F}\rangle$. --- \cite{gor-hae:j:compression-part-ii}
\end{quote}

Although it is not explicit in their argument, $\rho$ is simply a sentence whose provability in \mpl{} they are trying to decide.
Unfortunately, several notions in the proof are not properly defined. In order to analyze the above proof, we interpret those terms as faithfully as possible.
We assume that ``short'' means ``of height polynomial in $|\rho|$''
and that the ``guess'' is a nondeterministic guess by a nondeterministic 
polynomial-time machine. 
We also assume that ``small'' means ``of size polynomial in $|\rho|$.''
In this critique, we do not cover the definitions ``deduction frame,'' ``\textit{fst},''
``subdeduction,'' and so on. Rather, we refer interested readers to the original papers if they wish to have those definitions. However, the gist of the argument is as follows:
Nondeterministically guess a deduction of polynomial height (in $|\rho|$)
and deterministically compress it to one of polynomial size (in $|\rho|$). Finally
guess another deduction of size polynomial in $|\rho|$ and use both that newly guessed deduction and the compressed one to verify that $\rho$ is provable in \mpl.
Unfortunately, per Section~\ref{s:error}, 
there is no established polynomial upper bound on the size of proofs (i.e., deductions)
in the \mpl, and so, it's entirely possible that
no nondeterministic polynomial-time machine can guess a deduction for an
arbitrary (and provable in \mpl) $\rho$.
We thus observe a ripple effect, having its source in 
an earlier paper by Gordeev and Haeusler~\cite{gor-hae:j:compression-part-i}, that directly impacts
the central result of~\cite{gor-hae:j:compression-part-ii}.

\section{Conclusion}
Gordeev and Haeusler's \cite{gor-hae:j:compression-part-i} modifications to Hudelmaier's minimal propositional logic result in \LM{} and \NM{}, both of which are implicational systems that are not complete with respect to minimal propositional logic. Completeness is required to prove a polynomial bound on the size of all proofs of tautologies of minimal propositional logic, and extending \LM{} and \NM{} to be complete invalidates numerous lemmas and theorems that are required for the final result. The invalid result of \cite{gor-hae:j:compression-part-i}, which claims that any tautology of \NM{} has a DAG-like deduction of polynomial-bounded size, directly impacts their later paper, which uses the polynomial bound on proofs of minimal propositional logic to argue that $\np = \pspace$.

\section*{Acknowledgements}
We thank Ian Clingerman, Lane A. Hemaspaandra, and Quan Luu for their helpful comments on previous versions of this article. Additionally, we thank David E. Narv\'{a}ez for helping us get this project underway and into its current state. Any remaining errors are the responsibility of the authors.

\bibliographystyle{alpha}
\bibliography{main}

\end{document}